\documentclass[aps,prd,nofootinbib,showpacs,floatfix,twocolumn]{revtex4}

\usepackage{bm}
\usepackage{latexsym}
\usepackage{dcolumn}
\usepackage{amsfonts,amssymb,amsmath}
\usepackage{graphicx,epsfig}
\usepackage{psfrag}
\usepackage{subfigure}
\usepackage{feynmf}
\usepackage{rotating}
\usepackage{hyperref}
\usepackage{color}

\hypersetup{
    bookmarks=true,         
    unicode=false,          
    pdftoolbar=true,        
    pdfmenubar=true,        
    pdffitwindow=false,     
    pdfstartview={FitH},    
    pdftitle={My title},    
    pdfauthor={Author},     
    pdfsubject={Subject},   
    pdfcreator={Creator},   
    pdfproducer={Producer}, 
    pdfkeywords={keyword1} {key2} {key3}, 
    pdfnewwindow=true,      
    colorlinks=true,       
    linkcolor=red,          
    citecolor=cyan,        
    filecolor=magenta,      
    urlcolor=green,           
    linktocpage=true
}

\newcommand{\ee}{\mathrm{e}}

\newcommand{\ie}{\textsl{i.e.~}}

\def\beq{\begin{equation}}
\def\efq{\end{equation}}
\def\br{\begin{eqnarray}}
\def\er{\end{eqnarray}}
\def\benu{\begin{enumerate}}
\def\efnu{\end{enumerate}}
\def\nn{\nonumber}

\def\l{\left}
\def\r{\right}
\def\cR{{\cal R}}
\def\d{{\rm d}}
\def\f{\frac}

\def\ei{\eta_{\rm i}}
\def\ee{\eta_{\rm e}}
\def\ef{\eta_{\rm f}}
\def\te{t_{\rm e}}

\def\ae{a_{\rm e}}

\def\Ne{N_{\rm e}}
\def\Nf{N_{\rm f}}
\def\He{H_{\rm e}}

\def\vka{{\bm k}_{1}}
\def\vkb{{\bm k}_{2}}
\def\vkc{{\bm k}_{3}}
\def\ka{{\bm k}_{1}}
\def\kb{{\bm k}_{2}}
\def\kc{{\bm k}_{3}}

\def\cG{{\cal G}}
\def\cR{{\cal R}}
\def\cB{{\cal B}}
\def\fnl{f_{_{\rm NL}}}
\def\taunl{\tau_{_{\rm NL}}}
\def\Mp{M_{_{\rm Pl}}}

\begin{document}

\title{The scalar bi-spectrum during preheating in single 
field inflationary models}

\author{Dhiraj Kumar Hazra} \email{dhiraj@hri.res.in}
\affiliation{Harish-Chandra Research Institute, Chhatnag Road,
Jhunsi, Allahabad~211019, India}

\author{J\'er\^ome Martin} \email{jmartin@iap.fr}
\affiliation{Institut d'Astrophysique de Paris, UMR7095-CNRS,
  Universit\'e Pierre et Marie Curie, 98bis boulevard Arago, 75014
  Paris, France}

\author{L.~Sriramkumar} \email{sriram@physics.iitm.ac.in}
\affiliation{Department of Physics, Indian Institute of Technology Madras, 
Chennai~600036, India}

\begin{abstract}
  In single field inflationary models, preheating refers to the phase
  that immediately follows inflation, but precedes the epoch of
  reheating. During this phase, the inflaton typically oscillates at
  the bottom of its potential and gradually transfers its energy to
  radiation. At the same time, the amplitude of the fields coupled to
  the inflaton may undergo parametric resonance and, as a consequence,
  explosive particle production can take place. A priori, these
  phenomena could lead to an amplification of the super-Hubble scale
  curvature perturbations which, in turn, would modify the standard
  inflationary predictions. However, remarkably, it has been shown
  that, although the Mukhanov-Sasaki variable does undergo narrow
  parametric instability during preheating, the amplitude of the
  corresponding super-Hubble curvature perturbations remain
  constant. Therefore, in single field models, metric preheating does
  not affect the power spectrum of the large scale perturbations.  In
  this article, we investigate the corresponding effect on the scalar
  bi-spectrum. Using the Maldacena's formalism, we analytically show
  that, for modes of cosmological interest, the contributions to the
  scalar bi-spectrum as the curvature perturbations evolve on
  super-Hubble scales during preheating is completely
  negligible. Specifically, we illustrate that, certain terms in the
  third order action governing the curvature perturbations which may
  naively be expected to contribute significantly are exactly canceled
  by other contributions to the bi-spectrum. We corroborate selected
  analytical results by numerical investigations. We conclude with a
  brief discussion of the results we have obtained.
\end{abstract}

\pacs{98.80.Cq}

\maketitle


\section{Introduction}\label{sec:intro}

Inflation is currently considered the most attractive paradigm for
explaining the extent of homogeneity of the observable universe.  In
addition, the paradigm also provides a well motivated causal mechanism
for the generation of perturbations in the early universe (see any of
the following
texts~\cite{Kolb:1990aa,dodelson:2003,mukhanov2005physical,weinberg2008cosmology,durrer2008cosmic,peter2005cosmologie,lyth2009cosmology}
or one of the following
reviews~\cite{Kodama:1985bj,Mukhanov:1990me,Lidsey:1995np,Lyth:1998xn,Bassett:2005xm,Riotto:2002yw,Kinney:2009vz,Baumann:2009ds,Sriramkumar:2009kg,Martin:2003bt,Martin:2004um,Martin:2007bw,Linde:2007fr}).
In the standard picture, at the end of the inflationary epoch, the
inflaton, which is coupled to the other fields of the standard model,
decays into relativistic particles thereby transferring its energy to
radiation (see, for instance,
Refs.~\cite{Martin:2004um,Bassett:2005xm} and also
Refs.~\cite{Turner:1983he,Albrecht:1982mp,Martin:2006rs,Martin:2010kz}).
These decay products are then expected to
thermalize~\cite{Podolsky:2005bw} in order for the radiation dominated
epoch corresponding to the conventional hot big bang model to start.

\par 

In many models, inflation is terminated when the scalar field has
rolled down close to a minimum of the potential. Thereafter, the
scalar field usually oscillates at the bottom of the potential with an
ever decreasing amplitude.  There exists a period during this regime,
immediately after inflation but prior to the epoch of reheating, a
phase that is often referred to as
preheating~\cite{Traschen:1990sw,Shtanov:1994ce,Kofman:1997yn}.
During this brief phase, as in the inflationary era, the scalar field
continues to remain the dominant source that drives the expansion of
the universe.  Though the modes of cosmological interest
(corresponding to comoving wavenumbers $k$ such that, say, $10^{-4}< k
< 1\; {\rm Mpc}^{-1}$) are well outside the Hubble radius during this
phase, the conventional super-Hubble solutions to the curvature
perturbations that are applicable during inflation do not a priori
hold at this stage.  In fact, careful analysis is required to evolve
these modes during the phase of preheating.  However, despite the
subtle effects that need to be accounted for, it can be shown that, in
single field inflationary models, the amplitude of the curvature
perturbations and, hence, the amplitude as well as the shape of the
scalar power spectrum associated with the scales of cosmological
interest remain unaffected by the process of preheating (for the
original effort, see Ref.~\cite{Finelli:1998bu}; for more recent
discussions, see
Refs.~\cite{Jedamzik:2010dq,Jedamzik:2010hq,Easther:2010mr}).

\par

Over the last decade, there has been a tremendous theoretical interest
in understanding the extent of the non-Gaussianities that are
generated during
inflation~\cite{Gangui:1993tt,Gangui:1994yr,Gangui:1999vg,Gangui:2000gf,Maldacena:2002vr,Seery:2005wm,Chen:2005fe,Chen:2006nt,Langlois:2008wt,Langlois:2008qf,Chen:2010xka,Chen:2008wn,Chen:2006xjb,Hotchkiss:2009pj,Hannestad:2009yx,Flauger:2010ja,Adshead:2011bw,Adshead:2011jq,Martin:2011sn,Hazra:2012yn,Gangui:2002qc,Holman:2007na,Xue:2008mk,Meerburg:2009ys,Chen:2010bka}.
Simultaneously, there has been a constant effort to arrive at
increasingly tighter constraints on the dimensionless non-Gaussianity
parameter $\fnl$ that is often used to characterize the amplitude of
the reduced scalar bi-spectrum (viz. a suitable ratio of the scalar
bi-spectrum to the corresponding power spectrum) from the available
Cosmic Microwave Background (CMB)
data~\cite{Komatsu:2001rj,Komatsu:2003iq,Babich:2004yc,Liguori:2005rj,Hikage:2006fe,Fergusson:2006pr,Yadav:2007rk,Creminelli:2006gc,Yadav:2007yy,Hikage:2008gy,Rudjord:2009mh,Smith:2009jr,Smidt:2010ra,Fergusson:2010dm,Liguori:2010hx,Yadav:2010fz,Komatsu:2010hc,Larson:2010gs,Komatsu:2010fb}.
For instance, it has been theoretically established that slow roll
inflation driven by the canonical scalar field typically leads to
rather small values of $\fnl$ (of the order of the first slow roll
parameter)~\cite{Gangui:1993tt,Gangui:1994yr,Gangui:1999vg,Maldacena:2002vr}.
In contrast, though a Gaussian primordial perturbation lies well
within $2$-$\sigma$, the mean values of $\fnl$ from the CMB
observations seem to indicate a significant amount of
non-Gaussianity~\cite{Komatsu:2001rj,Komatsu:2003iq,Babich:2004yc,Liguori:2005rj,Hikage:2006fe,Fergusson:2006pr,Yadav:2007rk,Creminelli:2006gc,Yadav:2007yy,Hikage:2008gy,Rudjord:2009mh,Smith:2009jr,Smidt:2010ra,Fergusson:2010dm,Liguori:2010hx,Yadav:2010fz,Komatsu:2010hc,Larson:2010gs,Komatsu:2010fb}. Such
large levels for the parameter $\fnl$ can be generated when one
considers non-canonical scalar
fields~\cite{Seery:2005wm,Chen:2005fe,Chen:2006nt,Langlois:2008wt,Langlois:2008qf}
or when there exist deviations from slow roll
inflation~\cite{Chen:2008wn,Chen:2006xjb,Hotchkiss:2009pj,Hannestad:2009yx,Flauger:2010ja,Adshead:2011bw,Adshead:2011jq,Martin:2011sn,Hazra:2012yn}.

\par

We mentioned above that, in single field models, the epoch of
preheating does not affect the curvature perturbations and the scalar
power spectrum generated during inflation on cosmologically relevant
scales.  Note that, the scalar power spectrum is essentially
determined by the amplitude of the curvature perturbation.  Whereas,
as we shall discuss, the scalar bi-spectrum generated during inflation
involves integrals over the curvature perturbations as well as the
slow roll
parameters~\cite{Maldacena:2002vr,Seery:2005wm,Chen:2005fe,Chen:2006nt,Langlois:2008wt,Langlois:2008qf,Chen:2008wn,Chen:2006xjb,Hotchkiss:2009pj,Hannestad:2009yx,Flauger:2010ja,Adshead:2011bw,Adshead:2011jq,Martin:2011sn,Hazra:2012yn}.
If indeed deviations from slow roll inflation can result in high
levels of non-Gaussianity, then, naively, one may imagine that the
termination of inflation and the regime of preheating---both of which
involve large values for the slow roll parameters---can also lead to
large non-Gaussianities.  In other words, it may seem that the epoch
of preheating can contribute significantly to the scalar bi-spectrum.
In this work, we shall investigate the contributions to the scalar
bi-spectrum during preheating in single field inflationary models.
Remarkably, though the epoch of preheating actually amplifies specific
contributions to the bi-spectrum, as we shall illustrate, certain
cancellations arise that leave the total bi-spectrum generated during
inflation virtually unaltered.

\par

This paper is organized as follows.  In the following section, we
shall highlight the essential aspects of preheating in single field
inflationary models.  In particular, we shall discuss the behavior of
the scalar field as well as the scale factor, when a canonical scalar
field is oscillating at the bottom of an inflationary potential which
behaves quadratically near its minimum.  We shall also discuss a few
important points concerning the evolution of the curvature
perturbation on super-Hubble scales during preheating.  In
Sec.~\ref{sec:sbs}, after defining the bi-spectrum, we shall outline
the various contributions to the bi-spectrum in the Maldacena
formalism.  In Sec.~\ref{sec:sbs-cdp}, we shall evaluate the
contributions to the bi-spectrum for super-Hubble modes as the scalar
field oscillates in the quadratic potential, and show that the total
contribution during this epoch proves to be insignificant for these
modes.  We shall also support certain analytical results with the
corresponding numerical computations.  Finally, in Sec.~\ref{sec:d},
we shall conclude with a brief summary and outlook.

\par

A couple of words on our notation are in order at this stage of our
discussion.  We shall work with units such that $c=\hbar=1$ and shall
set $\Mp^2 =(8\, \pi\, G)^{-1}$.  An overdot and an overprime shall
denote differentiation with respect to the cosmic time $t$ and the
conformal time $\eta$, respectively.  Also, $N$ shall represent the
number of e-folds.  Moreover, double angular brackets shall denote
averaging over the oscillations during preheating.

 
\section{Behavior of the background and the large scale 
perturbations during preheating}

In this section, we shall discuss the behavior of the background and
the evolution of the curvature perturbation on super-Hubble scales
during preheating.  We shall consider a model involving the canonical
scalar field and assume that the inflationary potential behaves
quadratically around its minimum.

 
\subsection{Background evolution about a quadratic minimum}

\begin{figure*}[t]
\includegraphics[width=.45\textwidth,height=.325\textwidth]{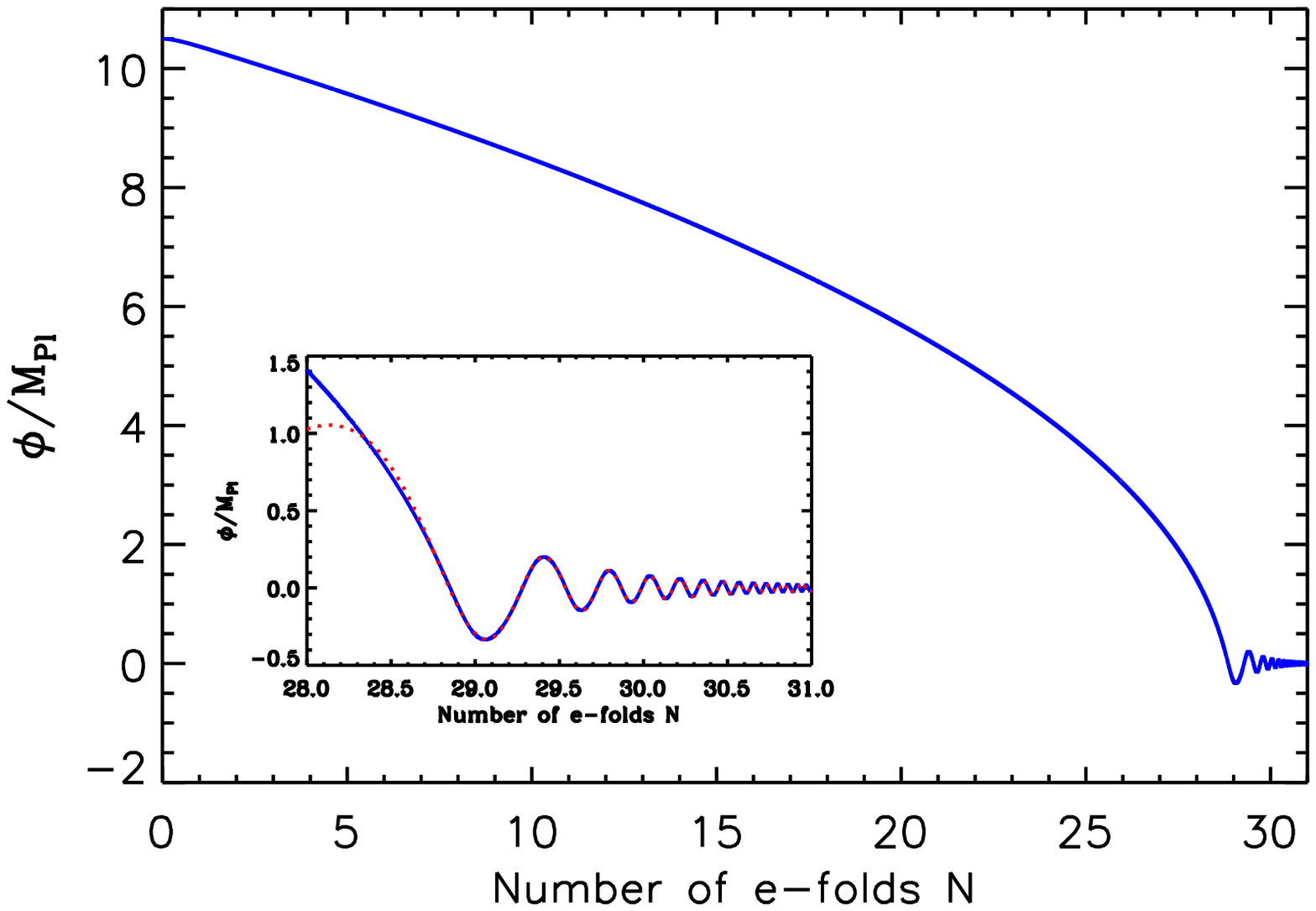}
\includegraphics[width=.45\textwidth,height=.325\textwidth]{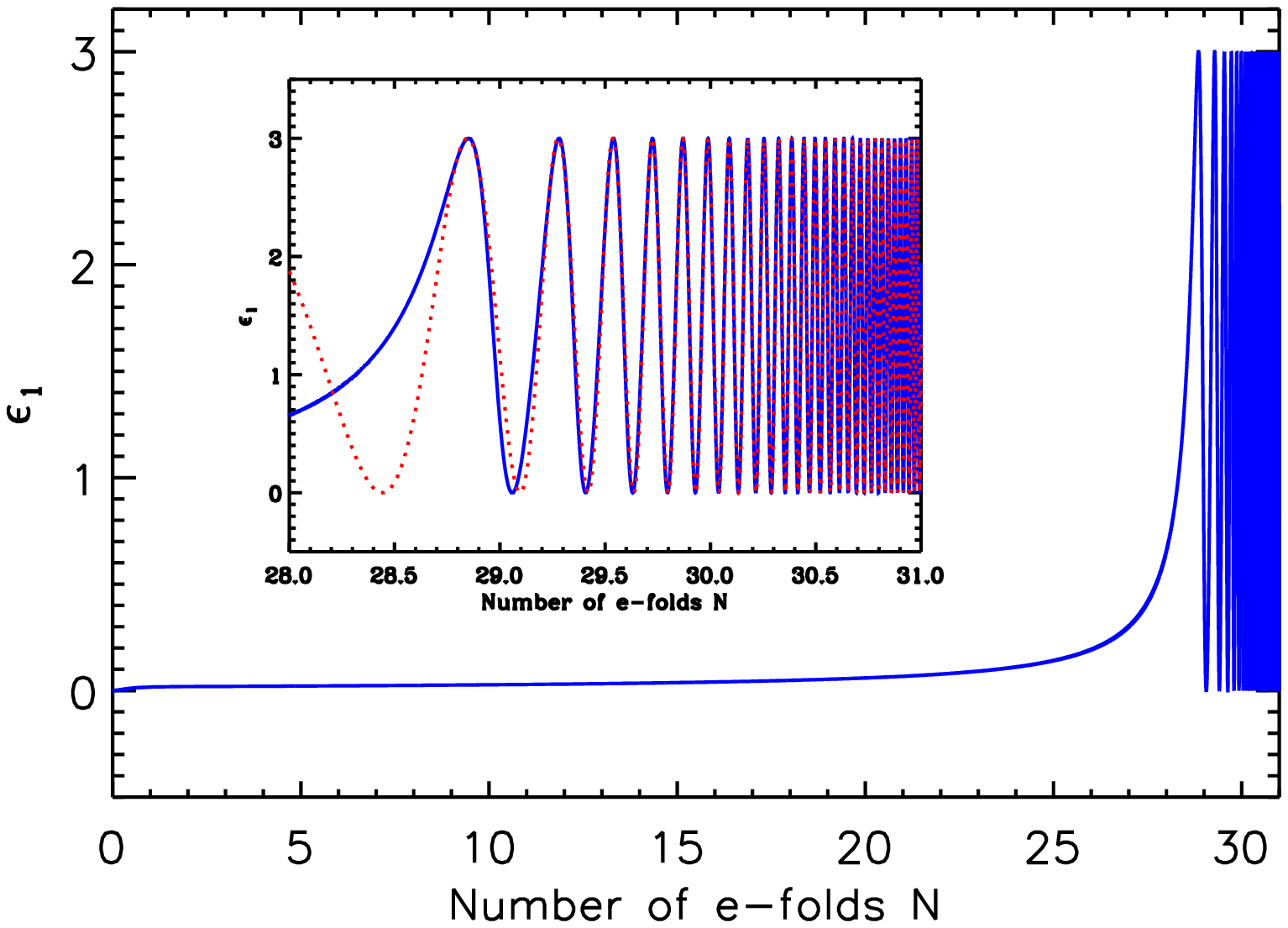}
\caption{The behavior of the scalar field (left panel) and the
  evolution of the first slow roll parameter $\epsilon_1$ (right
  panel) during the epochs of inflation and preheating have been
  plotted as a function of the number of e-folds for the case of the
  archetypical chaotic inflationary model described by the quadratic
  potential.  The blue curves denote the numerical results, while the
  dotted red curves in the insets represent the analytical results
  given by Eqs.~(\ref{eq:phi-prh}) and~(\ref{eq:e1-d-prh}) that are
  applicable during preheating. The analytical results evidently match
  the numerical ones quite well. Note that, for the choice parameters
  and initial conditions that we have worked with, $\epsilon_1$ turns
  unity at the e-fold of $\Ne \simeq 28.3$, indicating the termination
  of inflation at the point.  The fact that the field oscillates with
  a smaller and smaller amplitude once inflation has ended is clear
  from the inset (in the figure on the left panel). We should mention
  that we have worked with a smaller range of e-folds just for
  convenience.}\label{fig:fe-e1}
\end{figure*}

Consider a canonical scalar field $\phi$ that is governed by the
quadratic potential $V(\phi)=m^2\,\phi^2/2$ near its minimum.  It is
well known that, in such cases, slow roll inflation can be realized if
the field starts sufficiently far away from the minimum, with suitably
small values for its
velocity~\cite{Kolb:1990aa,dodelson:2003,mukhanov2005physical,weinberg2008cosmology,durrer2008cosmic,peter2005cosmologie,lyth2009cosmology,Kodama:1985bj,Mukhanov:1990me,Lidsey:1995np,Lyth:1998xn,Bassett:2005xm,Riotto:2002yw,Kinney:2009vz,Baumann:2009ds,Sriramkumar:2009kg,Martin:2003bt,Martin:2004um,Martin:2007bw,Linde:2007fr}.
Provided the initial conditions fall in the basin of the inflationary
attractor, the number of e-folds of inflation achieved largely depends
only on the initial value of the field, and inflation ends as the
field nears the bottom of the potential.  In fact, according to the
slow roll approximation, in an inflationary potential that consists of
no terms other than the above-mentioned quadratic one, inflation gets
terminated as the field crosses $\phi_{\rm e}=\sqrt{2}\; \Mp$.
Thereafter, the scalar field oscillates about the minimum with a
constantly decreasing amplitude because of the friction caused due to
the expansion.
These behavior are clearly evident from
Fig.~\ref{fig:fe-e1}, where we have plotted the evolution of the
scalar field and the first slow roll parameter~$\epsilon_1$, arrived
at numerically, both during and immediately after inflation for the
quadratic potential.
We should emphasize here that focusing on single field models can be 
considered to be essentially equivalent to assuming that the coupling 
of the inflaton to other fields is suitably weak during preheating.
The weak coupling will allow the condensate to live sufficiently long 
for a few oscillations to take place about the minimum of the inflaton
potential.

\par

\begin{figure}[t]
\includegraphics[width=.45\textwidth,height=.325\textwidth]{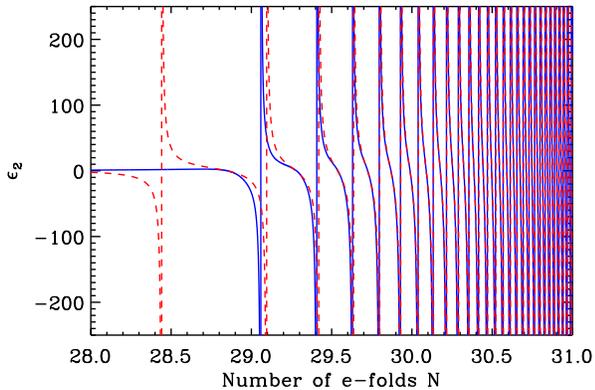}
\caption{The behavior of the second slow roll parameter $\epsilon_2$
  immediately after the termination of inflation has been plotted as a
  function of e-folds.  As in the previous figure, the blue curve
  represents the numerical result, while the dashed red curve denotes
  the analytical result during preheating
  [viz.~Eq.~(\ref{eq:e2-d-prh})].  Upon comparing this plot with the
  earlier plot of $\epsilon_1$, it is clear that $\epsilon_2$ diverges
  exactly at the turning points where $\epsilon_1$ vanishes, while
  $\epsilon_2$ itself vanishes whenever the field is at the bottom of
  the potential at which point $\epsilon_1$ attains its maximum
  value.}\label{fig:e2}
\end{figure}

Recall that, the first slow roll parameter is given by
\begin{equation}
\epsilon_1
=-\f{\dot H}{H^2}=\f{{\dot\phi}^2}{2\, H^2\, \Mp^2},\label{eq:e1}
\end{equation}
where $H\equiv \dot{a}/a$ is the Hubble parameter, with $a(t)$ being
the scale factor associated with the
Friedmann-Lemaitre-Robertson-Walker line-element. The second slow roll
parameter $\epsilon_2$ is defined in terms of the first one as
follows:
\begin{equation}
\epsilon_2\equiv \frac{1}{\epsilon_1}\,
\f{\d \epsilon_1}{\d N}
=\f{{\dot \epsilon}_1}{H\epsilon_1}.\label{eq:e2}
\end{equation}
Since the field is oscillating at the bottom of the potential post
inflation, $\dot\phi=0$ at the `turning points' and, hence,
$\epsilon_1=0$ at such instances. Also, when the field is at the
bottom of the potential, $V(\phi=0)=0$ so that $6\,H^2\,
\Mp^2={\dot\phi}^2$, corresponding to $\epsilon_1=3$.  Hence, we can
expect $\epsilon_1$ to oscillate between these two extreme values.
Moreover, the above expression for the second slow roll parameter in
terms of the first suggests that $\epsilon_2$ will vanish whenever
$\epsilon_1$ reaches the maximum value (\ie at the bottom of the
potential wherein ${\dot \epsilon}_1=0$), and that it will diverge at
the `turning points' wherein $\epsilon_1$ vanishes.  These behavior
too are indeed reflected in the plot of $\epsilon_1$ in
Fig.~\ref{fig:fe-e1} and, in Fig.~\ref{fig:e2}, where we have
plotted~$\epsilon_2$.

\par 

Let us now try to arrive at the complete behavior of the background
scalar field analytically.  During the phase of preheating, one finds
that the period of the oscillations (characterized by the inverse mass
in the case of the quadratic potential of our interest) is much
smaller than the time scales associated with the expansion, \ie the
inverse of the Hubble parameter $H^{-1}$.  In such a situation, to
understand the effects of the scalar field on the scale factor, one
can average over the oscillations and make use of the averaged energy
density of the scalar field to solve the first Friedmann equation.
One finds that, in the quadratic potential of our interest, the
expansion behaves as in a matter dominated epoch, with the scale
factor growing as $a(t)\propto t^{2/3}$, so that the Hubble parameter
behaves as
$H=2/(3\,t)$~\cite{Turner:1983he,Albrecht:1982mp,Martin:2004um,Bassett:2005xm}.

\par 

Therefore, at the time of preheating, the oscillating scalar field
satisfies the differential equation
\begin{equation}
\ddot\phi+\f{2}{t}\, \dot\phi+m^2\, \phi=0.
\end{equation}
The solution to this differential equation can be immediately written
down to be
\begin{equation}
\f{\phi(t)}{\Mp}= \f{\alpha}{m\,t}\; {\rm sin}\, (m\,t+\Delta),
\label{eq:phi-prh}
\end{equation}
where $\alpha$ is a dimensionless constant that we shall soon
determine, while $\Delta$ is an arbitrary phase chosen suitably to
match the transition from inflation to the matter dominated era.  The
`velocity' of the field is then given by
\begin{eqnarray}
\f{{\dot \phi}(t)}{\Mp}
&=& \f{\alpha}{t}\; 
\l[{\rm cos}\, (m\,t+\Delta)
-\frac{1}{m\,t}\; {\rm sin}\, (m\,t+\Delta)\r]\nn\\
&\simeq& \f{\alpha}{t}\; {\rm cos}\, (m\,t+\Delta),\label{eq:d-phi-prh}
\end{eqnarray}
where, in arriving at the second expression, for the sake of
consistency (\ie in having made use of the averaged energy density in
the first Friedmann equation to arrive at the scale factor), we have
ignored the second term involving $t^{-2}$.  Upon using the above
expressions for $\phi$ and ${\dot \phi}$ and the fact that
$H=2/(3\,t)$ in the first Friedmann equation, we obtain that
$\alpha=\sqrt{8/3}$.  Under these conditions, we find that the first
slow roll parameter simplifies to
\begin{equation}
\epsilon_1\simeq 3\; {\rm cos}^{2} (m\,t+\Delta)\label{eq:e1-d-prh}
\end{equation}
which, upon averaging, reduces to $3/2$, as is expected in a matter
dominated epoch. It is represented in Fig.~\ref{fig:fe-e1} (right
panel). On using the above result for $\epsilon_1$ in the
definition~(\ref{eq:e2}), the second slow roll parameter can be
obtained to be
\begin{equation}
\epsilon_2(t)\simeq -3\,m\,t\;\tan\,\l(m\,t+\Delta\r),\label{eq:e2-d-prh}
\end{equation}
which is illustrated in Fig.~\ref{fig:e2}.

\par

During preheating, we can write $a(t)=\ae\; (t/\te)^{2/3}$, where $\te$ and
$\ae$ denote the cosmic time and the scale factor at the end of inflation. 
We should mention here that, in addition to the phase $\Delta$, one requires 
the value of $\te$ in order to match the above analytical results for the 
scalar field and the slow roll parameters with the numerical results.
After setting $\te=\gamma\; [2/(3\, \He)]$, where $\He$ is the value of the
Hubble parameter at the end of inflation, we have chosen the quantity $\gamma$
and the phase $\Delta$ suitably so as to match the analytical expressions with 
the numerical results.  
It is clear from Figs.~\ref{fig:fe-e1} and~\ref{fig:e2} that the agreement 
between the numerical and the analytical results is indeed very good.
We should mention here that, for the results to match, we seem to require a 
$\gamma$ that is slightly larger than unity.
Actually, for the values of the parameters that we have worked with, we find
that we need to choose $\gamma$ to be about $1.18$ for the analytical results
to match the numerical ones.
The fact that $\gamma$ is not strictly unity need not come as a surprise.
After all, some time is bound to elapse as the universe makes the transition 
from an inflationary epoch to the behavior as in a matter dominated era.


\subsection{Evolution of the perturbations}

As is well known, the scalar perturbations are essentially described
by the curvature perturbation, say,~${\cal R}$.  The scalar power
spectrum ${\cal P}_{_{\rm S}}(k)$ can be defined through the two point
correlation function of the Fourier modes of the curvature
perturbation as
follows~\cite{Kodama:1985bj,Mukhanov:1990me,Lidsey:1995np,Lyth:1998xn,Sriramkumar:2009kg,Martin:2003bt,Martin:2004um,Martin:2007bw,Linde:2007fr}
\begin{equation}
\langle {\hat \cR}_{\bm k}\, {\hat \cR}_{\bm p}\rangle 
=\f{(2\, \pi)^2}{2\, k^3}\; {\cal P}_{_{\rm S}}(k)\;
\delta^{(3)}\l({\bm k}+{\bm p}\r).\label{eq:ps-d}
\end{equation}
The quantum operator ${\hat \cR}_{\bm k}$ in turn is decomposed based
on the modes $f_{\bm k}$ which are governed by the differential
equation
\begin{equation}
f_{\bm k}''+2\, \f{z'}{z}\, f_{\bm k}' + k^{2}\, f_{\bm k}=0,\label{eq:defk}
\end{equation}
where $z=\sqrt{2\,\epsilon_1}\, \Mp\, a$.
The scalar power spectrum can be expressed in terms of the modes $f_{\bm k}$ 
as
\begin{equation}
{\cal P}_{_{\rm S}}(k)
\equiv \frac{k^3}{2\,\pi ^2}
\left\vert f_{\bm k}\right\vert ^2\label{Pzeta}
\end{equation}
with the assumption that the spectrum is evaluated at suitably late
times when the modes are sufficiently outside the Hubble radius during
the era of
inflation~\cite{Kolb:1990aa,dodelson:2003,mukhanov2005physical,weinberg2008cosmology,peter2005cosmologie,Kodama:1985bj,Mukhanov:1990me,Lidsey:1995np,Lyth:1998xn,Sriramkumar:2009kg,Martin:2003bt,Martin:2004um,Martin:2007bw,Linde:2007fr}.

\par 

It often proves to be convenient to work in terms of the so-called
Mukhanov-Sasaki variable $v_{\bm k}$, which is defined as $v_{\bm k}
=z\,f_{\bm k}$. In terms of the variable $v_{\bm k}$, the above
equation of motion for $f_{\bm k}$ reduces to the following simple
form:
\begin{equation}
  v_{\bm k}''+\l(k^2-\f{z''}{z}\r)v_{\bm k}=0.\label{eq:de-vk}
\end{equation}
The initial conditions on the perturbations are imposed when the modes
are well inside the Hubble radius during inflation. The modes are
usually chosen to be in the Bunch-Davies vacuum, which amounts to
demanding that the Mukhanov-Sasaki variable $v_{\bm k}$ reduces to
following Minkowski-like positive frequency mode in the sub-Hubble
limit:
\begin{equation}
  \lim_{k/(a\,H)\rightarrow \infty} 
  v_{\bm k}=\f{1}{\sqrt{2\, k}}\; {\rm e}^{-i\,k\,\eta}.\label{eq:bd-ic}
\end{equation}

\par

As we shall see, the scalar bi-spectrum shall involve integrals over
the modes $f_{\bm k}$ and its derivative $f_{\bm k}'$ as well as the
slow roll parameters $\epsilon_1$, $\epsilon_2$ and the derivative
$\epsilon_2'$.  So, in order to analyze the effects on the bi-spectrum
due to preheating, it becomes imperative that, in addition to the
behavior of the slow roll parameters, we also understand the evolution
of the mode $f_{\bm k}$ and its derivative during this epoch.  We have
already studied the behavior of the first two slow roll parameters in
the previous sub-section. Therefore, our immediate aim will be to
understand the evolution of the curvature perturbations for scales of
cosmological interest during the preheating phase.

\par

Since the modes of cosmological interest are well outside the Hubble
radius [\ie $k/(a\,H)\ll 1$] at late times, we need to arrive at the
super-Hubble solution for the mode $f_{\bm k}$ or, equivalently, the
Mukhanov-Sasaki variable $v_{\bm k}$.  In a slow roll inflationary
regime, \ie when $\l(\epsilon_1,\epsilon_2,\epsilon_3\r) \ll 1$, the 
effective potential $z''/z$ that governs the evolution of $v_{\bm k}$
[cf.~Eq.~(\ref{eq:de-vk})] can be written as
\begin{equation}
\f{z''}{z}=a^2\,H^2\, 
\l[2+{\cal O}\l(\epsilon_1,\epsilon_2,\epsilon_3\r)\r]
\simeq 2\, a^2\, H^2.\label{eq:ep-srl}
\end{equation}
Due to this reason, during slow roll inflation, the super-Hubble
condition $k/(a\,H)\ll 1$ amounts to neglecting the $k^2$ term with
respect to the effective potential $z''/z$ in the differential
equation~(\ref{eq:de-vk}).  In such a case, it is straightforward to
show that the super-Hubble solution to $v_{\bm k}$ up to the order
$k^2$ can be expressed as
follows~\cite{Kolb:1990aa,dodelson:2003,mukhanov2005physical,weinberg2008cosmology,peter2005cosmologie,Kodama:1985bj,Mukhanov:1990me,Lidsey:1995np,Lyth:1998xn,Sriramkumar:2009kg,Martin:2003bt,Martin:2004um,Martin:2007bw,Linde:2007fr}:
\begin{eqnarray}
  v_{\bm k}(\eta) 
  &\simeq& A_{\bm k}\, z(\eta)\; \l[1-k^2\,\int^{\eta}
  \f{\d{\bar \eta}}{z^2({\bar \eta})}\,
  \int^{\bar \eta} \d{\tilde \eta}\;z^2({\tilde \eta})\r]\nn\\
  & &+\, B_{\bm k}\, z(\eta)\, \int ^{\eta}
  \f{\d{\bar \eta}}{z^2({\bar \eta})}\nn\\
  & &\times\,
  \biggl[1-k^2\,\int ^{\bar \eta}\d {\tilde \eta}\; z^2({\tilde \eta})\,
  \int^{\tilde \eta}\frac{\d{\breve \eta}}{z^2({\breve \eta})}\biggr],
\label{eq:sh-solution}
\end{eqnarray}
where $A_{\bm k}$ and $B_{\bm k}$ are $k$-dependent constants that are
determined by the Bunch-Davies initial condition~(\ref{eq:bd-ic})
imposed in the sub-Hubble limit.  As is well known, the first term
involving $A_{\bm k}$ represents the growing mode, while the second
containing $B_{\bm k}$ corresponds to the decaying mode.  However, it
is important to realize that, at the time of preheating, the effective
potential $z''/z$ is no longer given by the slow roll
expression~(\ref{eq:ep-srl}).  It is clear that the effective
potential will contain oscillatory functions and, hence, it can even
possibly vanish.  So, it is not a priori obvious that one can use the
same approach as in the inflationary epoch and simply ignore the $k^2$
term in the differential equation~(\ref{eq:de-vk}) for arriving at the
behavior of the super-Hubble modes.  Moreover, it is known that,
during the preheating phase, one has to deal with the resonant
behavior exhibited by the equation of motion under certain
conditions~\cite{Finelli:1998bu,Finelli:2000ya,Jedamzik:2010dq}.  As a
consequence, at this stage, it becomes necessary that we remain
cautious and analyze equation~(\ref{eq:de-vk}) more carefully.

\par

In order to study the perturbations during the preheating phase, it
proves to be more convenient to work in terms of cosmic time and use a
new rescaled variable ${\cal V}_{\bm k}$ that is related to the
Mukhanov-Sasaki variable as follows: ${\cal V}_{\bm k}\equiv a^{1/2}\,
v_{\bm k}$.  Then, one finds that Eq.~(\ref{eq:de-vk}) takes the
form~\cite{Finelli:1998bu,Jedamzik:2010dq}
\begin{eqnarray}
{\ddot {\cal V}}_{\bm k} 
&+& \Biggl[\frac{k^2}{a^2} +\frac{{\rm d}^2V}{{\rm d}\phi ^2}
+\f{3\,\dot{\phi}^2}{\Mp^2}
-\frac{\dot{\phi}^4}{2\,H^2\,\Mp^4}\nn\\ 
&+&\frac{3}{4\, \Mp^2}\left(\frac{\dot{\phi}^2}{2}-V\right)
+\frac{2\,\dot{\phi}}{H\, \Mp^2}\,\frac{{\rm d}V}{{\rm d}\phi}\Biggr]\;
{\cal V}_{\bm k}=0.\quad
\end{eqnarray}
Recall that, in the quadratic potential of our interest, soon after
inflation, the evolution of the scalar field $\phi(t)$ is given by
Eq.~(\ref{eq:phi-prh}). Using this solution and its
derivative~(\ref{eq:d-phi-prh}), it is then easy to show that, while
the third, fourth and the fifth terms within the square brackets in
the above differential equation decay as $a^{-3}$, the last term
decays more slowly as it scales as $a^{-3/2}$.  Upon retaining only
the first, second and the last terms and neglecting the others, one
arrives at an equation of the form
\begin{eqnarray}
  \frac{\d^2 {\cal V}_{\bm k}}{\d \sigma^2}
  &+&\Biggl[1+\frac{k^2}{m^2\,a^2}\nn\\
  &-&\f{4}{m\, \te}\,
  \l(\frac{\ae}{a}\r)^{3/2}\,
  \cos\,\l(2\,\sigma+2\,\Delta\r)\Biggr]\, {\cal V}_{\bm k}=0,
  \quad\label{eq:de-calVk}
\end{eqnarray}
where the new independent variable $\sigma$ is a dimensionless quantity
which we have defined to be $\sigma\equiv m\, t+\pi/4$.  We can rewrite the
above equation as
\begin{equation}
\f{\d^2 {\cal V}_{\bm k}}{\d \sigma^2} 
+\l[{\cal A}_k-2\,q\;
{\rm cos}\,\l(2\,\sigma+2\,\Delta\r)\r]\, {\cal V}_{\bm k} = 0,
\end{equation}
with ${\cal A}_k$ and $q$ being given by
\begin{eqnarray}
{\cal A}_k &=& 1+\frac{k^2}{m^2a^2},\\
\label{eq:q}
q &=& \frac{2}{m\,\te}\; \l(\frac{\ae}{a}\r)^{3/2},
\end{eqnarray} 
where, as we mentioned, $\te$ and $\ae$ denote the cosmic time and the
scale factor when inflation ends.  The above equation is similar in
form to the Mathieu equation (see, for instance,
Ref.~\cite{bender1978advanced}). The Mathieu equation possesses
unstable solutions that are known to grow rapidly when the values of
the parameters fall in certain domains known as the resonant bands.
As discussed in detail in Refs.~\cite{Kofman:1997yn,Jedamzik:2010dq},
since $q\ll 1$ in the situation of our interest, one falls in the
narrow resonance regime. In such a case, the first instability band is
delineated by the condition $1-q < {\cal A}_k <1+q$, which turns out
to be equivalent to the condition
\begin{equation}
0<\frac{k}{a}<\sqrt{3\,H\,m}.\label{eq:first-band}
\end{equation}
It should be emphasized here that the time evolution of the quantities
${\cal A}_k$ and $q$ are such that, once a mode has entered the
resonance band, it remains inside it during the entire oscillatory
phase.

\par

Note that, in Eq.~(\ref{eq:de-calVk}), we can neglect the term
involving $k^2$ provided $k^2/(m^2\, a^2)\ll 1$.  This condition can
be rewritten as
\begin{equation}
\left(\frac{k}{a\,H}\right)^2\,\frac{H^2}{m^2}\ll 1.
\label{eq:condition-1}
\end{equation}
On the other hand, the condition to fall in the first instability
band, viz. Eq.~(\ref{eq:first-band}), can be expressed
as~\cite{Jedamzik:2010dq}
\begin{equation}
\left(\frac{k}{a\,H}\right)^2\, \frac{H}{3\,m}\ll 1.
\label{eq:condition-2}
\end{equation}
Given that, $H<m$ immediately after inflation, it is evident that the
first of the above two conditions will be satisfied if the second is.
In other words, being in the first instability band implies that one
can indeed neglect the $k^2$ term in Eq.~(\ref{eq:de-calVk}).  But,
clearly, this is completely equivalent to ignoring the $k^2$ term in
the original equation~(\ref{eq:de-vk}).  Therefore, we can conclude
that, provided we fall in the first instability band (which is the
case for the range of modes and parameters of our interest), it is
perfectly valid to work with the super-Hubble
solution~(\ref{eq:sh-solution}) even during the preheating phase.

\par

The above conclusion can also supported by the following arguments. As
discussed in Ref.~\cite{Jedamzik:2010dq}, in the first instability
band, the Floquet index is given by $\mu =q/2$. In such a case, the
mode ${\cal V}_{\bm k}$ behaves as ${\cal V}_{\bm k}\propto 
{\rm e}^{\mu\, \sigma}$. However, in the situation of our interest, since 
we have a time dependent Floquet index, the corresponding solution can be
written as
\begin{equation}
{\cal V}_{\bm k}
\propto \exp\, \l(\int\,\mu \, \d \sigma\,\r)\propto a^{3/2}
\end{equation}
which, in turn, implies that $v_{\bm k}={\cal V}_{\bm
  k}/a^{1/2}\propto a$.  Further, since, $f_{\bm k} =v_{\bm k}/z$ and
$z\propto a$ during preheating (\ie if one makes use of the fact that
$\epsilon_1=3/2$ on the average), we arrive at the result that $f_{\bm
  k}$ remains a constant during this phase. This property is indeed
the well known behavior one obtains if one simply retains the very
first term of the growing mode in the super-Hubble
solution~(\ref{eq:sh-solution}).  But, it should be realized that the
above arguments also demonstrate another important point.  Note that
$f_{\bm k}$ is a constant not only for modes on super-Hubble scales
but, for all the modes (even those that remain in the sub-Hubble
domain), provided they fall in the resonance band.  This behavior can
possibly be attributed to the background. After all, it is common
knowledge that the amplitude of the curvature perturbations remain
constant on all scales in a matter dominated
era~\cite{Kolb:1990aa,dodelson:2003,mukhanov2005physical,weinberg2008cosmology,peter2005cosmologie,Kodama:1985bj,Mukhanov:1990me,Lidsey:1995np,Lyth:1998xn,Sriramkumar:2009kg,Martin:2003bt,Martin:2004um,Martin:2007bw,Linde:2007fr}.

\par

We have proven that, in the first instability band and on super-Hubble
scales, the solution~(\ref{eq:sh-solution}) is valid during
preheating.  Let us now analyze this solution in further detail.  In
particular, in order to check the extent of its validity, let us
compare the analytical estimate for the curvature perturbation with
the numerical solution. It is clear that the
solution~(\ref{eq:sh-solution}) leads to the following expression for
the growing mode:
\begin{eqnarray}
  f_{\bm k}(\eta) 
  \simeq A_{\bm k}\, \l[1-k^2\,
  \int ^{\eta} \frac{{\rm d}{\bar \eta}}{z^2({\bar \eta})}\,
  \int^{\bar \eta}
  {\rm d}{\tilde \eta}\; z^2({\tilde \eta})\right],
\label{eq:zetalargescale}
\end{eqnarray}
where we have retained the scale dependent correction for comparison
with the numerical result. Since it is the contribution due to the
growing mode that will prove to be dominant, we shall compare the
behavior of the above solution during preheating with the
corresponding numerical result.  We shall carry out the comparison for
a suitably small scale mode so that the second term in the above
expression is not completely insignificant.

\par

\begin{figure}[!t]
\begin{center}
\includegraphics[width=8.65cm]{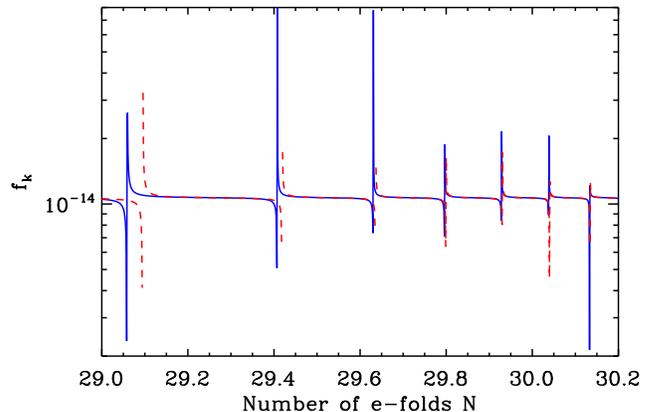}
\end{center}
\caption{The behavior of the curvature perturbation during preheating.
  The blue curve denotes the numerical result, while the dashed red
  curve represents the analytical solution~(\ref{eq:fk}).  We have
  chosen to work with a very small scale mode $k$ that leaves the
  Hubble radius at about two e-folds before the end of inflation.  We
  have made use of the same value of $\Delta$ as in the previous two
  figures and we have fixed $A_{\bm k}$ [cf. Eq.~(\ref{eq:fk})] by
  choosing it to be the numerical value of the curvature perturbation
  at a suitable time close to the end of inflation.  It is clear that
  the agreement between the analytical and the numerical results is
  quite good.}
\label{fig:R}
\end{figure}

We now need to evaluate the double integral in the above expression
for $f_{\bm k}$ during preheating.  Let us write
\begin{equation}
K(t)= \int^{\eta} \frac{{\rm d}{\bar \eta}}{z^2({\bar \eta})}\;
J({\bar \eta})
=\int ^{t} \frac{{\rm d}{\bar t}}{a({\bar t})\, z^2({\bar t})}\;
J({\bar t}),
\end{equation}
where 
\begin{equation}
J({\bar t})
= \int^{{\bar \eta}} {\rm d}{\tilde \eta}\; z^2({\tilde \eta})
=\int^{\bar t} \frac{{\rm d}{\tilde t}}{a({\tilde t})}\; 
z^2({\tilde t}).
\end{equation}
Upon making use of the matter dominated behavior of the scale factor
and the expression~(\ref{eq:e1-d-prh}) for $\epsilon_1$, we find that
the integral $J({\bar t})$ can be performed exactly. We obtain that
\begin{eqnarray}
J({\bar t})&=&\frac{3}{2}\,\Mp^2\, \ae\,\te\, 
\biggl[\f{6}{5}\,\l(\f{{\bar t}}{\te}\r)^{5/3}\nn\\
& &+\,{\rm e}^{2\,i\,\Delta}\,\l(-2\,i\,m\,\te\r)^{-5/3}\,
\gamma\l(\frac{5}{3},-2\,i\,m\,{\bar t}\r)\nn\\
& &+\,{\rm e}^{-2\,i\,\Delta}\,\l(2\,i\,m\,\te\r)^{-5/3}\,
\gamma\l(\frac{5}{3},2\,i\,m\,{\bar t}\r)+{\cal C}\biggr],\qquad
\label{eq:J}
\end{eqnarray}
where $\gamma(b,x)$ is the incomplete Gamma function (see, for
example, Refs.~\cite{Abramovitz:1970aa,Gradshteyn:1965aa}), while the
quantity ${\cal C}$ is a dimensionless constant of integration. Then, the
relation between the incomplete Gamma and the Gamma functions allows
us to express the function $K(t)$ as follows:
\begin{widetext}
\begin{eqnarray}
K(t) &\simeq& \f{\te}{4\,\ae^2}
\int^{t} \frac{{\rm d}{\bar t}}{\cos^2(m\,{\bar t}+\Delta)}
\Biggl[\f{6}{5} \l(\f{\te}{\bar t}\r)^{1/3}
+{\rm e}^{2\,i\,\Delta}\l(-2\,i\,m\,\te\r)^{-5/3}\,
\Gamma\l(\f{5}{3}\r) \l(\f{\te}{\bar t}\r)^{2}
-{\rm e}^{2\,i\,\l(m\,{\bar t}+\Delta\r)}
\l(-2\,i\,m\,\te\r)^{-1} \l(\f{\te}{\bar t}\r)^{4/3}\nonumber\\ 
& &+\,{\rm e}^{-2\,i\,\Delta}\,\l(2\,i\,m\,\te\r)^{-5/3}\,
\Gamma\l(\f{5}{3}\r)\, \l(\f{\te}{\bar t}\r)^{2}
- {\rm e}^{-2\,i\,\l(m\,{\bar t}+\Delta\r)}\,
\l(2\,i\,m\,\te\r)^{-1}\, \l(\f{\te}{\bar t}\r)^{4/3} 
+{\cal C}\,\l(\f{\te}{\bar t}\r)^2\Biggr]
\nonumber \\
&\simeq & \f{3\, \te^{4/3}}{10\, \ae^2}
\int^{t} \frac{{\rm d}{\bar t}\;\; 
{\bar t}^{-1/3}}{\cos^2(m\,{\bar t}+\Delta)}+\cdots .
\end{eqnarray}
\end{widetext}
In arriving at the final equality, we have used the asymptotic
property of the incomplete Gamma
function~\cite{Abramovitz:1970aa,Gradshteyn:1965aa} and have retained
only the dominant term in inverse power of $m\,t$.  The final
expression above can be integrated by parts to arrive at
\begin{eqnarray}
K(t)&\simeq& \f{3\, \te^{4/3}}{10\, m\,\ae^2}\; t^{-1/3}\,
\tan\, (m\,t+\Delta)\nn\\
&+ &\,\f{\te^{4/3}}{10\, m\,\ae^2}\;
\int^{t} {\rm d}{\bar t}\;
{\bar t}^{-4/3}\, \tan\,(m\,{\bar t}+\Delta).
\end{eqnarray}
The second term containing the integral in this expression is of the
order of the other terms that we have already neglected and, hence, it
too can be ignored. As a result, the growing mode of the curvature
perturbation can be written as
\begin{eqnarray}
f_{\bm k} 
&\simeq& A_{\bm k}\, \Biggl[1-\f{3}{10}\,
\frac{k^2\,\te^{4/3}}{\ae^2\,m\,t^{1/3}}\,\tan\,(m\,t+\Delta)\Biggr]\nn\\
&=& A_{\bm k}\,\Biggl[1-\frac{1}{5}\, \l(\frac{k}{a\,H}\r)^2\,
\frac{H}{m}\, \tan\,(m\,t+\Delta)\Biggr],\label{eq:fk}
\end{eqnarray}
in perfect agreement with the result that has been obtained recently
in the literature~\cite{Easther:2010mr}. It is evident from the above
expression that the evolution of the curvature perturbation will
contain sharp spikes during preheating, a feature that is clearly
visible in Fig.~\ref{fig:R} wherein we have plotted the above
analytical expression as well as the corresponding numerical result
(in this context, also see Fig.~4 in Ref.~\cite{Jedamzik:2010dq} where
the spikes are also clearly visible).

\par

It is important that we make a couple of remarks concerning the
appearance of the spikes in the evolution of the curvature
perturbation.  Firstly, as the spikes are encountered both
analytically and numerically, evidently, they are not artifacts of the
adopted approach. Secondly, one may fear that the perturbation theory
would break down as soon as one encounters a spike, which indicates a
rather large value for the perturbation variable of interest.  We
believe that such issues could possibly be avoided when one couples
the inflaton to radiation, as is needed to reheat the universe.



\section{The scalar bi-spectrum in the Maldacena formalism}
\label{sec:sbs}

Now that we have understood the behavior of the large scale modes at
the time of preheating, let us turn to investigate the effects of
preheating on the bi-spectrum.  In this section, we shall quickly
sketch the various contributions to the bi-spectrum in the Maldacena
formalism~\cite{Maldacena:2002vr}.

\par 

The scalar bi-spectrum $\cB_{_{\rm S}}(\vka,\vkb,\vkc)$ is defined in
terms of the three point correlation function of the Fourier modes of
the curvature perturbation $\cR$ as
follows~\cite{Larson:2010gs,Komatsu:2010fb}:
\begin{eqnarray}
\langle {\hat \cR}_{\vka}\, 
{\hat \cR}_{\vkb}\, {\hat \cR}_{\vkc}\rangle 
&=&\l(2\,\pi\r)^3\; \cB_{_{\rm S}}(\vka,\vkb,\vkc)\nn\\
& &\times\,\delta^{(3)}\l(\vka+\vkb+\vkc\r).\label{eq:bs-d}
\end{eqnarray}
For convenience, we shall set
\begin{equation}
G(\vka,\vkb,\vkc)
=(2\,\pi)^{9/2}\, \cB_{_{\rm S}}(\vka,\vkb,\vkc).
\end{equation}
In the Maldacena formalism to calculate the
bi-spectrum~\cite{Maldacena:2002vr}, the quantity $G(\vka,\vkb,\vkc)$
can be expressed
as~\cite{Seery:2005wm,Chen:2005fe,Chen:2010xka,Martin:2011sn,Hazra:2012yn}
\begin{widetext}
\begin{eqnarray}
G(\vka,\vkb,\vkc)
&\equiv & \sum_{C=1}^{7}\; G_{_{C}}(\vka,\vkb,\vkc)
\nonumber \\
&\equiv & \Mp^2\; \sum_{C=1}^{6}\; 
\biggl[f_{\ka}(\ef)\, f_{\kb}(\ef)\,f_{\kc}(\ef)\, 
\cG_{_{C}}(\vka,\vkb,\vkc)
+\,f_{\ka}^{\ast}(\ef)\, f_{\kb}^{\ast}(\ef)\,
f_{\kc}^{\ast}(\ef)\, \cG_{_{C}}^{\ast}(\vka,\vkb,\vkc)\biggr]\nonumber \\ & &
+\, G_{7}(\vka,\vkb,\vkc),
\end{eqnarray}
where $\ef$ denotes the final time when the bi-spectrum is to be
evaluated.  The quantities $\cG_{_{C}}(\vka,\vkb,\vkc)$ with $C
=(1,6)$ are described by the
integrals~\cite{Seery:2005wm,Chen:2005fe,Chen:2010xka,Martin:2011sn,Hazra:2012yn}
\begin{eqnarray}
\cG_{1}(\vka,\vkb,\vkc)
&=& 2\,i\,\int_{\ei}^{\ef} \d\eta\; a^2\, 
\epsilon_{1}^2\; \bigl(f_{\ka}^{\ast}\,f_{\kb}'^{\ast}\,
f_{\kc}'^{\ast}+\,{\rm two~permutations}\bigr),\label{eq:cG1}\\
\cG_{2}(\vka,\vkb,\vkc)
&=&-\,2\,i\;\l(\vka\cdot \vkb + {\rm two~permutations}\r)
\,\int_{\ei}^{\ef} \d\eta\; a^2\, 
\epsilon_{1}^2\, f_{\ka}^{\ast}\,f_{\kb}^{\ast}\,
f_{\kc}^{\ast},\label{eq:cG2}\\
\cG_{3}(\vka,\vkb,\vkc)
&=&-\,2\,i\,\int_{\ei}^{\ef} \d\eta\; a^2\,
\epsilon_{1}^2
\, \Biggl[\l(\f{\vka\cdot\vkb}{\kb^{2}}\r)\,
f_{\ka}^{\ast}\,f_{\kb}'^{\ast}\, f_{\kc}'^{\ast}
+\, {\rm five~permutations}\Biggr],\label{eq:cG3}\\
\cG_{4}(\vka,\vkb,\vkc)
&=&i\,\int_{\ei}^{\ef} \d\eta\; a^2\,\epsilon_{1}\,
\epsilon_{2}'\,
\,\bigl(f_{\ka}^{\ast}\,f_{\kb}^{\ast}\,
f_{\kc}'^{\ast}+{\rm two~permutations}\bigr),\label{eq:cG4}\\
\cG_{5}(\vka,\vkb,\vkc)
&=&\frac{i}{2}\,\int_{\ei}^{\ef} \d\eta\;
a^2\, \epsilon_{1}^{3}
\,\Biggl[\l(\f{\vka\cdot\vkb}{\kb^{2}}\r)\,
f_{\ka}^{\ast}\,f_{\kb}'^{\ast}\, f_{\kc}'^{\ast}
+\, {\rm five~permutations}\Biggr],\label{eq:cG5}\\
\cG_{6}(\vka,\vkb,\vkc) 
&=&\frac{i}{2}\,\int_{\ei}^{\ef} \d\eta\; a^2\, 
\epsilon_{1}^{3}
\,\Biggl\{\l[\f{\ka^{2}\,\l(\vkb\cdot\vkc\r)}{\kb^{2}\,\kc^{2}}\r]\, 
f_{\ka}^{\ast}\, f_{\kb}'^{\ast}\, f_{\kc}'^{\ast}
+\, {\rm two~permutations}\Biggr\},\label{eq:cG6}
\end{eqnarray}
\end{widetext}
where $\ei$ denotes the time when the modes $f_{\bm k}$ are well
inside the Hubble radius during inflation.  The additional, seventh
term $G_{7}(\vka,\vkb,\vkc)$ arises due to a field redefinition, and
its contribution to $G(\vka,\vkb,\vkc)$ is found to be
\begin{eqnarray}
G_{7}(\vka,\vkb,\vkc)
&=&\frac{\epsilon_{2}(\ef)}{2}\,
\biggl[\vert f_{\kb}(\ef)\vert^{2}\, 
\vert f_{\kc}(\ef)\vert^{2}\nn\\
&+& \, {\rm two~permutations}\biggr].\label{eq:G7}
\end{eqnarray} 


\section{The contributions to the scalar bi-spectrum during 
preheating}
\label{sec:sbs-cdp}

Our goal now is to determine the different contributions to the
bi-spectrum due to the epoch of preheating for modes of cosmological
interest. Since we know the behavior of the slow roll parameters and
the large scale modes during preheating, it is clear from the above
expressions that, it is simply a matter of being able to evaluate the
integrals involved.  As we shall illustrate, it turns out to be
possible to actually evaluate these integrals explicitly and thereby
arrive at the contributions to the bi-spectrum due to preheating.

\par 

During preheating, for the large scale modes (\ie for $k\ll a\,H$),
the contribution to $f_{\bm k}$ is dominated by the constant growing
mode [viz. the very first term in Eq.~(\ref{eq:sh-solution})] so that
one has
\begin{equation}
f_{\bm k}\simeq A_{\bm k}.\label{eq:fk-shs-d-prh}
\end{equation}
As the corresponding derivative trivially vanishes, at the same order
in $k$, the leading non-zero contribution to the quantity $f_{\bm k}'$
is determined by the decaying mode, which is given by
\begin{equation}
f_{\bm k}\simeq B_{\bm k}\, \int ^{\eta}
\f{\d{\bar \eta}}{z^2({\bar \eta})}
\end{equation}
and, hence,
\begin{equation}
f_{\bm k}'\simeq \f{B_{\bm k}}{z^2}
=\f{{\bar B}_{\bm k}}{a^2\, \epsilon_1},\label{eq:fkp-shs-d-prh}
\end{equation}
where we have set ${\bar B}_{\bm k}=B_{\bm k}/(2\, \Mp^2)$.

\par 

Let us now turn to discuss the various contributions to the bi-spectrum 
as the background and the modes evolve during preheating.


\subsection{The fourth and the seventh terms}

As we had mentioned in the introductory section, it has been realized
that departures from slow roll during inflation can lead to
significant
non-Gaussianities~\cite{Chen:2008wn,Chen:2006xjb,Hotchkiss:2009pj,Hannestad:2009yx,Flauger:2010ja,Adshead:2011bw,Adshead:2011jq,Martin:2011sn,Hazra:2012yn}.
In such situations, it has been repeatedly noticed that it is the
fourth term, viz. $G_4(\vka,\vkb,\vkc)$, that contributes the most to
the bi-spectrum because it involves the derivative of the second slow
roll parameter $\epsilon_2$.  Since $\epsilon_2$ grows extremely large
during preheating [see Eq.~(\ref{eq:e2-d-prh}) as well as
Fig.~\ref{fig:e2}], it is natural to expect that the fourth term will
contribute significantly to the bi-spectrum at the time of preheating.
So, we shall first investigate the contribution due to
$G_4(\vka,\vkb,\vkc)$.  As we shall see, it indeed proves to be large
during preheating.  However, interestingly, as we shall illustrate,
this large contribution is canceled by a similar contribution due to
the seventh term $G_7(\vka,\vkb,\vkc)$ which also involves the second
slow roll parameter~$\epsilon_2$.

\par

Upon using the super-Hubble behavior~(\ref{eq:fk-shs-d-prh}) 
and~(\ref{eq:fkp-shs-d-prh}) of the mode $f_{\bm k}$ and its 
derivative in the expression~(\ref{eq:cG4}), we obtain that
\begin{eqnarray}
\cG_{4}(\vka,\vkb,\vkc)
&\simeq& i\, \l(A_{{\bm k}_{1}}^{\ast}\, A_{{\bm k}_{2}}^{\ast}\, 
{\bar B}_{{\bm k}_3}^{\ast}\, +~{\rm two~permutations}\r)\nn\\
& &\times\, \int_{\ee}^{\ef} \d\eta\; \epsilon_{2}',
\end{eqnarray}
where $\ee$ denotes the time at which inflation ends.
The above expression can be trivially integrated to yield
\begin{eqnarray}
\cG_{4}(\vka,\vkb,\vkc)
&\simeq& i\, \l(A_{{\bm k}_{1}}^{\ast}\, A_{{\bm k}_{2}}^{\ast}\, 
{\bar B}_{{\bm k}_3}^{\ast}\, +~{\rm two~permutations}\r)\nn\\
& &\times\, \l[\epsilon_2(\ef)-\epsilon_2(\ee)\r],
\end{eqnarray}
so that the corresponding contribution to the bi-spectrum can be
expressed as
\begin{eqnarray}
G_{4}(\vka,\vkb,\vkc)
&\simeq& i\, \Mp^2\, \l[\epsilon_2(\ef)-\epsilon_2(\ee)\r]\nn\\
& &\times\,\biggl[\vert A_{{\bm k}_1}\vert^2\, \vert A_{{\bm k}_2}\vert^2
\l(A_{{\bm k}_3}\, {\bar B}_{{\bm k}_3}^{\ast}-A_{{\bm k}_3}^{\ast}\, 
{\bar B}_{{\bm k}_3}\r)\nn\\
& &+\,~{\rm two~permutations}\biggr].\label{eq:G4-d-prh}
\end{eqnarray}
Since this expression is proportional to $\epsilon_2$, it suggests 
that preheating may contribute substantially to the bi-spectrum.
But, as we shall soon show, this large contribution is canceled by a 
similar contribution from the seventh term that arises due to the 
field redefinition [cf.~Eq.~(\ref{eq:G7})].

\par

Now, consider the Wronskian
\begin{equation}
{\cal W}=f_{\bm k}\, f_{\bm k}'^{\ast}-f_{\bm k}^{\ast}\, f_{\bm k}'.
\end{equation}
Upon using the equation of motion~(\ref{eq:defk}) for $f_{\bm k}$, one
can show that, ${\cal W}=W/z^{2}$, where $W$ is a constant.  It is
important to note that this result is valid on all scales, even in the
sub-Hubble limit during inflation.  In this limit, as we had
mentioned, the modes $v_{\bm k}$ satisfy the Bunch-Davies initial
condition~(\ref{eq:bd-ic}).  On making use of this sub-Hubble behavior
in the above definition of the Wronskian~${\cal W}$, one obtains that
$W=i$.  In the super-Hubble limit, we have, on using the corresponding
solution~(\ref{eq:fk-shs-d-prh}) and its
derivative~(\ref{eq:fkp-shs-d-prh}),
\begin{equation}
{\cal W}= \f{2\,\Mp^2}{z^{2}}\, \,
\l(A_{\bm k}\, {\bar B}_{\bm k}^{\ast}\
-A_{\bm k}^{\ast}\, {\bar B}_{\bm k}\r)=\f{i}{z^2}.
\end{equation}
Therefore, we obtain that  
\begin{equation}
\l(A_{\bm k}\, {\bar B}_{\bm k}^{\ast}\
-A_{\bm k}^{\ast}\, {\bar B}_{\bm k}\r)
=\f{i}{2\, \Mp^2},
\end{equation}
and, hence, the expression~(\ref{eq:G4-d-prh}) for $G_{4}(\vka,\vkb,\vkc)$ 
simplifies to
\begin{eqnarray}
G_{4}(\vka,\vkb,\vkc)
&\simeq& -\f{1}{2}\, \l[\epsilon_2(\ef)-\epsilon_2(\ee)\r]\nn\\
& &\times\,\biggl[\vert A_{{\bm k}_1}\vert^2\, \vert A_{{\bm k}_2}\vert^2\nn\\
& &+\,~{\rm two~permutations}\biggr].
\end{eqnarray}
Note that the first of these terms [involving $\epsilon_2(\ef)$] 
{\it exactly}\/ cancels the contribution $G_7(\vka,\vkb,\vkc)$
[cf.~Eq.~(\ref{eq:G7})] that arises due to the field redefinition
(with $f_{\bm k}$ set to $A_{\bm k}$), leaving behind only the
contributions generated during inflation.

\par

Before we go on to discuss the behavior of the other contributions, we
should emphasize here that the above result for the fourth and the
seventh terms applies to all single field models.  It is important to
appreciate the fact that we have made no assumptions whatsoever about
the inflationary potential in arriving at the above conclusion. However, 
one should keep in mind that, regarding its behavior near the minima,  
we have made use of the fact that the potential can be approximated by a 
parabola. Indeed, it is with this explicit form that we have been able 
to identify a solution to the Mukhanov-Sasaki equation that leads to a 
constant curvature perturbation.


\subsection{The second term}

We shall now turn to the second term $G_2(\vka,\vkb,\vkc)$, which in
certain situations is known to be as large as the fourth term when
there exist periods of fast roll~\cite{Martin:2011sn,Hazra:2012yn}.
During preheating, we have
\begin{eqnarray}
  \cG_{2}(\vka,\vkb,\vkc)
  &=&-2\,i\;\l(\vka\cdot \vkb + {\rm two~permutations}\r)\nn\\
  & &\times\,A_{{\bm k}_1}^{\ast}\,A_{{\bm k}_2}^{\ast}\,A_{{\bm k}_3}^{\ast}\;
  I_2(\ef,\ee),
\end{eqnarray}
where $I_2(\ef,\ee)$ denotes the integral
\begin{equation}
I_2(\ef,\ee)=\int_{\ee}^{\ef} \d\eta\; a^2\, \epsilon_{1}^2,
\end{equation}
so that the corresponding contribution to the bi-spectrum is given by
\begin{eqnarray}
G_2(\vka,\vkb,\vkc)
\!&=&\!-2\, i\,\Mp^2\,\l(\vka\cdot \vkb + {\rm two~permutations}\r)\nn\\
& &\times\, \vert A_{{\bm k}_1}\vert^2\, 
\vert A_{{\bm k}_2}\vert^2\, 
\vert A_{{\bm k}_3}\vert^2\nn\\
& &\times\,\l[I_2(\ef,\ee)-I_2^{\ast}(\ef,\ee)\r],
\end{eqnarray}
which identically vanishes since $I_2$ is real. Needless to add, this
implies that the second term does not contribute to the bi-spectrum
during preheating. Again we should emphasize the fact that, as in the
case of the fourth and the seventh terms, this result holds good for
any inflationary model provided it can be approximated by a parabola
in the vicinity of its minimum.

\par 

\begin{figure}[!htb]
\begin{center}
\includegraphics[width=8.65cm]{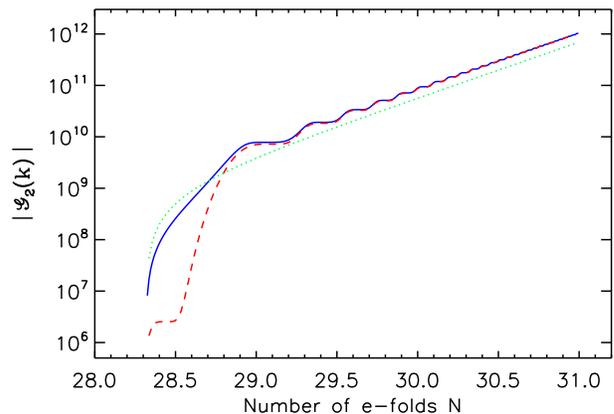}
\end{center}
\caption{The behavior of the quantity $\cG_{2}(\vka,\vkb,\vkc)$ in the
  equilateral limit, \ie when $k_1=k_2=k_3=k$, for a mode that leaves
  the Hubble radius at about $20$ e-folds before the end of inflation.
  The blue curve represents the numerical result.  The dashed red
  curve denotes the analytical result arrived at using the
  integral~(\ref{eq:I2-ef}) and with the same choice of $\Delta$ as in
  the earlier figures.  The dotted green curve corresponds to the
  integral~(\ref{eq:I2-ef-wo}) obtained when the oscillations have
  been ignored.  As in the previous figure, the value of $A_{\bm k}$
  has been fixed by choosing it to be the numerical value of the
  curvature perturbation on super-Hubble scales.  Needless to add, the
  match between the analytical and the numerical results is
  excellent.}
\label{fig:cG2-pi}
\end{figure}

In order to check that our assumptions and approximations are indeed
valid, let us now estimate the quantity $\cG_{2}(\vka,\vkb,\vkc)$
analytically during preheating for the case of the quadratic potential
and compare with the corresponding numerical result. In such a case,
the integral $I_2(\ef,\ee)$ can be carried out along similar lines to
the integral $J({\bar t})$ that we had evaluated earlier
[cf.~Eq.~(\ref{eq:J})].  We find that it can be expressed in terms of
the incomplete Gamma function $\gamma(b,x)$ as follows:
\begin{widetext}
\begin{eqnarray}
I_2(\ef,\ee)&=&\f{9\, \ae\,\te}{16}\; (m\, \te)^{-5/3}\,
\Biggl(\f{18}{5}\, (m\, \te)^{5/3}\,
\l[{\rm e}^{5\, (\Nf-\Ne)/2}-1\r]\nn\\
& &+\,4\; (-2\,i)^{-5/3}\; {\rm e}^{2\,i\,\Delta}\;
\l\{\gamma\l[\f{5}{3},-2\,i\,m\,t_{\rm e}\, {\rm e}^{3\, (\Nf-\Ne)/2}\r]
-\gamma\l(\f{5}{3},-2\,i\,m\,t_{\rm e}\r)\r\}\nn\\
& &+\,4\; (2\,i)^{-5/3}\, {\rm e}^{-2\,i\,\Delta}\;
\l\{\gamma\l[\f{5}{3},2\,i\,m\,t_{\rm e}\, {\rm e}^{3\, (\Nf-\Ne)/2}\r]
-\gamma\l(\f{5}{3},2\,i\,m\,t_{\rm e}\r)\r\}\nn\\
& &+\,(-4\,i)^{-5/3}\;{\rm e}^{4\,i\,\Delta}\;
\l\{\gamma\l[\f{5}{3},-4\,i\,m\,t_{\rm e}\, {\rm e}^{3\, (\Nf-\Ne)/2}\r]
-\gamma\l(\f{5}{3},-4\,i\,m\,t_{\rm e}\r)\r\}\nn\\
& &+\, (4\,i)^{-5/3}\;{\rm e}^{-4\,i\,\Delta}\;
\l\{\gamma\l[\f{5}{3},4\,i\,m\,t_{\rm e}\, {\rm e}^{3\, (\Nf-\Ne)/2}\r]
-\gamma\l(\f{5}{3},4\,i\,m\,t_{\rm e}\r)\r\}\Biggr).\label{eq:I2-ef}
\end{eqnarray}
\end{widetext}
On the other hand, had we ignored the oscillations during preheating, 
and assumed that the background behavior is exactly the same as in a 
matter dominated era, then, since, $\langle\!\langle\epsilon_1\rangle
\!\rangle=3/2$, the quantity $\langle\!\langle I_2(\ef,\ee)\rangle
\!\rangle$ can be trivially evaluated to yield
\begin{equation}
\langle\!\langle I_2(\ef,\ee)\rangle\!\rangle
=\f{27\, \ae\,\te}{20}\; \l[{\rm e}^{5\, (\Nf-\Ne)/2}-1\r].
\label{eq:I2-ef-wo}
\end{equation}
We have plotted the quantity $\cG_2(\vka,\vkb,\vkc)$ in the
equilateral limit, \ie when $k_1=k_2=k_3=k$, corresponding to the
analytical expressions~(\ref{eq:I2-ef}) and~(\ref{eq:I2-ef-wo}) as
well as the numerical result as a function of upper limit $\Nf$ during
preheating in Fig.~\ref{fig:cG2-pi}. The agreement between the
analytical and the numerical results is indeed striking.

\subsection{The remaining terms}

In this sub-section, we shall analytically compute the contributions
due to the remaining terms, viz. the first, third, fifth and the
sixth.  Notice that, the first term $G_1(\vka,\vkb,\vkc)$ and the
third term $G_3(\vka,\vkb,\vkc)$ involve the same integrals.
Therefore, these two contributions to the bi-spectrum can be clubbed
together.  Similarly, the fifth and the sixth terms,
viz. $G_5(\vka,\vkb,\vkc)$ and $G_6(\vka,\vkb,\vkc)$, also contain
integrals of the same type, and hence their contributions too can be
combined.

\par 

On making use of the super-Hubble behavior~(\ref{eq:fk-shs-d-prh})
and~(\ref{eq:fkp-shs-d-prh}) of the mode $f_{\bm k}$ and its
derivative, we obtain that, during preheating,
\begin{eqnarray}
\cG_{1}(\vka,\vkb,\vkc)
\!&\simeq\!& 2\,i \l(A_{{\bm k}_{1}}^{\ast}\, 
{\bar B}_{{\bm k}_{2}}^{\ast}\, 
{\bar B}_{{\bm k}_3}^{\ast}\, +~{\rm two~permutations}\r)\nn\\
& &\times\, I_{13}(\ef,\ee)\qquad
\end{eqnarray}
and
\begin{eqnarray}
\cG_{3}(\vka,\vkb,\vkc)
&\simeq& -\,2\,i\, \biggl[\l(\f{\vka\cdot\vkb}{\kb^{2}}\r)\,
A_{{\bm k}_{1}}^{\ast}\, {\bar B}_{{\bm k}_{2}}^{\ast}\, 
{\bar B}_{{\bm k}_3}^{\ast}\nn\\ 
& &+~{\rm five~permutations}\biggr]\; I_{13}(\ef,\ee).\qquad\;\;
\end{eqnarray}
The quantity $I_{13}(\ef,\ee)$ represents the integral
\begin{equation}
I_{13}(\ef,\ee)= \int_{\ee}^{\ef} \f{\d\eta}{a^2},
\end{equation}
which can be trivially carried out during preheating to yield
\begin{equation}
I_{13}(\ef,\ee)
=\f{\te}{\ae^3}\, \l[1-{\rm e}^{-3\,(\Nf-\Ne)/2}\r].
\end{equation}
From these results, we find that the contribution to the bi-spectrum 
due to the first and the third terms can be written as
\begin{widetext}
\begin{eqnarray}
G_1(\vka,\vkb,\vkc)+G_3(\vka,\vkb,\vkc)
&=& 2\,i\,\Mp^2\,I_{13}(\ef,\ee)\,
\Biggl[\l(1-\f{\vka\cdot\vkb}{\kb^{2}}
-\f{\vka\cdot\vkc}{\kc^{2}}\r)\vert A_{{\bm k}_{1}}\vert^2\nonumber\\ 
& &\times\, \l(A_{{\bm k}_{2}}{\bar B}_{{\bm k}_{2}}^{\ast}A_{{\bm k}_{3}}
{\bar B}_{{\bm k}_3}^{\ast}-A_{{\bm k}_{2}}^{\ast} {\bar B}_{{\bm k}_{2}}
A_{{\bm k}_{3}}^{\ast}{\bar B}_{{\bm k}_3}\r)
+~{\rm two~permutations}\Biggr].
\end{eqnarray}
Since the second term in the above expression for $I_{13}(\ef,\ee)$
dies quickly with growing $\Nf$, the corresponding contribution to 
the bi-spectrum proves to be negligible.

\par

The contributions due to the fifth and the sixth terms during
preheating can be arrived at in a similar fashion.  We obtain that
\begin{eqnarray}
G_5(\vka,\vkb,\vkc)+G_6(\vka,\vkb,\vkc)
&=&\f{i\,\Mp^2}{2}\,\,I_{56}(\ef,\ee)\,
\Biggl\{\l[\f{\vka\cdot\vkb}{\kb^{2}}
+\f{\vka\cdot\vkc}{\kc^{2}}
+\f{\ka^{2}\,\l(\vkb\cdot\vkc\r)}{\kb^{2}\,\kc^{2}}\r]
\vert A_{{\bm k}_{1}}\vert^2
\bigl(A_{{\bm k}_{2}}\, {\bar B}_{{\bm k}_{2}}^{\ast}\, A_{{\bm k}_{3}}\,
{\bar B}_{{\bm k}_3}^{\ast}\nonumber \\ 
& &-\,A_{{\bm k}_{2}}^{\ast}\, {\bar B}_{{\bm k}_{2}}\,
A_{{\bm k}_{3}}^{\ast}\, {\bar B}_{{\bm k}_3}\bigr)
+~{\rm two~permutations}\Biggr\},
\end{eqnarray}
with $I_{56}(\ef,\ee)$ denoting the integral
\begin{equation}
I_{56}(\ef,\ee)
=\int_{\ee}^{\ef} \f{\d\eta}{a^2}\, \epsilon_1.
\end{equation}
This integral too can be evaluated rather easily to arrive at
the following expression:
\begin{eqnarray}
I_{56}(\ef,\ee)
&=&\f{3\,\te}{\ae^3}\; 
\Biggl(\cos^2\left(m\,\te+\Delta\right)
- {\rm e}^{-3\,(\Nf-\Ne)/2}\,
\cos^2\left[m\,\te \, {\rm e}^{3\,(\Nf-\Ne)/2}+\Delta\right]\nn\\
& &+\,m\,\te\, \cos\left(2\,\Delta\right)\,
\left\{{\rm Si}\l(2\,m\,\te\r)
-{\rm Si}\left[2\,m\,\te\, {\rm e}^{3\,(\Nf-\Ne)/2}\right]\right\}\nn\\ 
& &+\,m\,\te\, \sin\left(2\,\Delta\right)
\left\{{\rm Ci}\left(2\,m\,\te\right)
-{\rm Ci}\left[2\,m\,\te\, {\rm e}^{3(\Nf-\Ne)/2}\right]\right\}\Biggr),
\end{eqnarray}
\end{widetext} 
where ${\rm Si}(x)$ and ${\rm Ci}(x)$ are the sine and the cosine integral
functions~\cite{Abramovitz:1970aa,Gradshteyn:1965aa}. 
And, had we ignored the oscillations, we would have arrived at 
\begin{equation}
\langle\!\langle I_{56}(\ef,\ee)\rangle\!\rangle
=\f{3\,\te}{2\,\ae^3}\; \l[1-{\rm e}^{-3\,(\Nf-\Ne)/2}\r],
\end{equation}
which is of the same order as $I_{13}(\ef,\ee)$, and hence completely
negligible as we had discussed.


\subsection{The contribution to $\fnl$ during preheating}

Let us now actually estimate the extent of the contribution to the 
non-Gaussianity parameter $\fnl$ during preheating.
Since the contributions due to the combination of the fourth plus the 
seventh and the second term completely vanish at late times, the 
non-zero contribution to the bi-spectrum during preheating is 
determined by the first, third, fifth and the sixth terms.
Note that, if one ignores the oscillations post-inflation, then one 
has $I_{56}(\ef,\ee)=3\,I_{13}(\ef,\ee)/2$. 
In such a situation, we find that the non-trivial contributions lead to 
the following bi-spectrum in the simpler case of the equilateral limit:
\begin{eqnarray}
G_{\rm eq}(k) &=&\f{69\, i\,\Mp^2}{8}\,\,I_{13}(\ef,\ee)\,
\vert A_{{\bm k}}\vert^2\nn\\
& &\times\;\l(A_{\bm k}^2\, {\bar B}_{\bm k}^{\ast}{}^2\,
- A_{\bm k}^{\ast}{}^2\, {\bar B}_{\bm k}^2\r).\nn\\
\end{eqnarray}
In the equilateral limit, the non-Gaussianity parameter $\fnl$ is given
by 
\begin{equation}
\fnl^{\rm eq}(k)=-\f{10}{9}\, \f{1}{\l(2\,\pi\r)^4}\, 
\f{k^6\, G_{\rm eq}(k)}{{\cal P}_{_{\rm S}}^{2}(k)},
\end{equation}
where ${\cal P}_{_{\rm S}}(k)$ is the power spectrum defined in 
Eq.~(\ref{eq:ps-d}).
Upon making use of the fact that $f_{\bm k}\simeq A_{\bm k}$ at late times, 
we then obtain the contribution to $\fnl^{\rm eq}$ during preheating to be
\begin{eqnarray}
\label{eq:fnlreheat}
\fnl^{\rm eq}(k)
&\simeq& -\f{115\, i\,\Mp^2}{48}\,\,I_{13}(\ef,\ee)\nn\\
& &\times\, \l(\f{A_{\bm k}^2\, {\bar B}_{\bm k}^{\ast}{}^2\,
- A_{\bm k}^{\ast}{}^2\, {\bar B}_{\bm k}^2}{\vert A_{\bm k}\vert^2}\r).
\end{eqnarray}

\par

In order to explicitly calculate the parameter $\fnl$, we need to first 
specify the inflationary scenario. 
We shall choose to work with power law inflation because it permits exact 
calculations, and it can also mimic slow roll inflation. 
During power law inflation, the scale factor can be written as $a(\eta)=a_1\, 
(\eta/\eta_1)^{\beta+1}$, where $a_1$ and $\eta_1$ are constants, while 
$\beta$ is a free index. 
It is useful to note that, in such a case, the first slow roll parameter 
is a constant and is given by $\epsilon_1=(\beta +2)/(\beta +1)$. 
The current constraints on the scalar spectral index suggest that $\beta 
\lesssim -2$, which implies that the corresponding scale factor is close 
to that of de Sitter. 

\par

In power law inflation, the exact solution to Eq.~(\ref{eq:de-vk}) can be
expressed in terms of the Bessel function $J_{\nu}(x)$ as follows:
\begin{equation}
\label{eq:solbessel}
v_{\bm k}(\eta)=\sqrt{-k\,\eta}\; \left[C_{\bm k}\,J_{\nu}(-k\,\eta)
+D_{\bm k}\,J_{-\nu}(-k\,\eta)\right],
\end{equation}
where $\nu=(\beta+1/2)$, and the quantities $C_{\bm k}$ and $D_{\bm k}$ 
are constants that are determined by the initial conditions.
Upon demanding that the above solution satisfies the Bunch-Davies initial
condition~(\ref{eq:bd-ic}), one obtains that
\begin{eqnarray}
C_{\bm k}&=&-D_{\bm k}\; {\rm e}^{-i\,\pi\,(\beta +1/2)}, \\ 
D_{\bm k}&=& \sqrt{\frac{\pi}{k}}\; 
\frac{{\rm e}^{i\,\pi\,\beta/2}}{2\,\cos\,(\pi\,\beta)}.
\end{eqnarray}
One can confirm that the super-Hubble limit of the solution~(\ref{eq:solbessel}) 
above indeed reproduces Eq.~(\ref{eq:sh-solution}) exactly. Moreover, the limit 
also allows us to arrive at the constants $A_{\bm k}$ and $B_{\bm k}$, which are 
found to be
\begin{eqnarray}
A_{\bm k}&=& \frac{2^{-(\beta+1/2)}}{\Gamma(\beta+3/2)}\;
\frac{(-k\,\eta_1)^{\beta +1}}{\sqrt{2\,\epsilon_1}\,a_1\,\Mp}\, C_{\bm k},\\
B_{\bm k}&=& -\frac{(2\,\beta+1)\;2^{\beta +1/2}}{\Gamma(-\beta+1/2)}\;
\f{\sqrt{2\,\epsilon_1}\,a_1\, \Mp}{\eta_1}\nn\\
& &\times\, (-k\,\eta_1)^{-\beta}\, D_{\bm k}.
\end{eqnarray}
Then, upon inserting the above expressions for the quantities $A_{\bm k}$ and 
$B_{\bm k}$ in Eq.~(\ref{eq:fnlreheat}), we find that
\begin{eqnarray}
\fnl^{\rm eq}(k)&=& \frac{115\,\epsilon_1}{288\,\pi}\; 
\Gamma^2\l(\beta +\f{1}{2}\r)\,
2^{2\beta+1}\, \l(2\,\beta+1\r)^2\nn\\ 
& &\times\,\sin\,(2\,\pi\,\beta)\,\left\vert \beta+1\right\vert^{-2\,(\beta+1)}\nn\\
& &\times\,\l[1-{\rm e}^{-3\,(\Nf-\Ne)/2}\r]\,
\left(\frac{k}{\ae H_{\rm e}}\right)^{-(2\,\beta+1)}.\qquad
\end{eqnarray}
This expression can also be rewritten in terms of the parameters describing 
the post-inflationary evolution. 
We obtain that
\begin{widetext}
\begin{eqnarray}
\fnl^{\rm eq}(k) &=& \frac{115\, \epsilon_1}{288\,\pi}\;\Gamma^2\l(\beta +\f{1}{2}\r)\,
2^{2\beta+1}\, \l(2\,\beta+1\r)^2\,\sin\,(2\,\pi\,\beta)\,
\left\vert \beta+1\right\vert^{-2\,(\beta+1)}\, 
\l[1-{\rm e}^{-3\,(\Nf-\Ne)/2}\r]\nn\\
& & \times\, \left[\left(\frac{\pi^2\,g_*}{30}\right)^{-1/4}\,
\left(1+z_{\rm eq}\right)^{1/4}\,\frac{\rho_{\rm cri}^{1/4}}{T_{\rm rh}}
\right]^{-(2\,\beta +1)}\, \left(\frac{k}{a_0\, H_0}\right)^{-(2\,\beta+1)},
\end{eqnarray}
\end{widetext}
where $g_*$ denotes the effective number of relativistic degrees of 
freedom at reheating, $T_{\rm rh}$ the reheating temperature and 
$z_{\rm eq}$ the redshift at the epoch of equality.
Also, $\rho_{\rm cri}$, $a_0$ and $H_0$ represent the critical energy 
density, the scale factor and the Hubble parameter today, respectively.
The above expression is mainly determined by the ratio 
$\rho_{\rm cri}^{1/4}/T_{\rm rh}$. 
For a model with $\beta \simeq -2$ and a reheating temperature of $T_{\rm rh}
\simeq 10^{10}\, \mbox{GeV}$, one obtains that $\fnl \sim 10^{-60}$ for 
the modes of cosmological interest (\ie for $k$ such that $k/a_0\simeq H_0$), 
a value which is completely unobservable. 
This confirms and quantifies our result that, in the case of single field inflation, 
the epoch of preheating does not alter the amplitude of the scalar bi-spectrum 
generated during inflation~\cite{Kohri:2009ac}.
However, it is worthwhile to add that, while the amplitude of the above
non-Gaussianity parameter $\fnl$ is small, it seems to be strongly scale
dependent.

\par

We believe that a couple of points require further emphasis 
at this stage of our discussion.
Recall that, to fall within the first instability band during preheating, 
the modes need to satisfy the condition~(\ref{eq:first-band}).
But, in order to neglect the term involving $k^2$ in the differential
equation~(\ref{eq:de-calVk}), the modes of interest are actually 
required to satisfy the condition~(\ref{eq:condition-1}).
As we have emphasized earlier, evidently, the condition~(\ref{eq:condition-1}) 
will be easily satisfied by the large scale modes that already lie within the 
instability band and thereby satisfying the condition~(\ref{eq:condition-2}).
Therefore, it is important to appreciate the fact that the conclusions we have 
arrived at above apply to all cosmologically relevant scales.


\section{Discussion}\label{sec:d}

In this work, we have analyzed the effects of preheating on the
primordial bi-spectrum in inflationary models involving a single
canonical scalar field.  We have illustrated that, certain
contributions to the bi-spectrum, such as those due to the combination
of the fourth and the seventh terms and that due to the second term,
vanish identically at late times. Further, assuming the inflationary
potential to be quadratic around its minimum, we have shown that the
remaining contributions to the bi-spectrum are completely
insignificant during the epoch of preheating when the scalar field is
oscillating at the bottom of the potential immediately after
inflation. It is important to appreciate the fact that the results 
we have arrived at apply to any single field inflationary potential that 
has a parabolic shape near the minimum.

\par 
 
A couple of other points also need to be stressed regarding the 
conclusions we have arrived at. The results we have
obtained supplement the earlier results wherein it has been shown that
the power spectrum generated during inflation remains unaffected
during the epoch of preheating (see, Ref.~\cite{Finelli:1998bu}; in
this context, also see Ref.~\cite{Liddle:1999hq}). Moreover, our
results are in support of earlier discussions which had pointed to the
fact that the contributions to the correlation functions at late times
will be insignificant if the interaction terms in the actions at the
cubic and the higher orders depend on either a time or a spatial
derivative of the curvature
perturbation~\cite{Weinberg:2005vy,Weinberg:2006ac,Weinberg:2008nf,Weinberg:2008si,Weinberg:2010wq}.

\par

Broadly, our effort needs to be extended in two different directions.
Firstly, it is important to confirm that the conclusions we have
arrived at hold true for potentials which behave differently, say,
quartically, near the minima. Further, the exercise needs to be
repeated for models involving non-canonical scalar 
fields~\cite{Lorenz:2008je, Lorenz:2008et,Lachapelle:2008sy}. 
In this context, it is worth mentioning that the generalization of the
conserved quantity ${\cal R}_{\bm k}$ in the Dirac-Born-Infeld case has 
been shown to stay constant in amplitude on scales larger than the sonic 
horizon, a property which allows us to propagate the spectrum from horizon 
exit till the beginning of the radiation dominated era~\cite{Lorenz:2008et}. 
Secondly, as we had mentioned, preheating is followed by an epoch of reheating 
when the energy from the inflaton is expected to be transferred to radiation. 
It will be interesting to examine the evolution of the bi-spectrum during
reheating. However, in order to achieve reheating, the scalar field
needs to be coupled to radiation. It is clear that the formalism for
evaluating the bi-spectrum involving just the inflaton is required to
be extended to a situation wherein radiation too is present and is
also coupled to the scalar field.  

\par

However, possibly, the most interesting direction opened up by our
work concerns multi-field inflation and associated
non-Gaussianities~\cite{Rigopoulos:2005ae,Seery:2005gb,Arroja:2008yy,Gao:2008dt,Mizuno:2009mv,RenauxPetel:2011uk}.
Unlike single field models wherein the curvature perturbation
associated with the large scale modes is conserved at late times, such
a behavior is not necessarily true in multi-field inflation.  When
many fields are present, the entropy (\ie the iso-curvature)
fluctuations can cause the evolution of curvature perturbations even
on super-Hubble scales.  Further, in the case of multi-field
inflation, it is known that the two-point correlation function can be
affected by metric preheating~\cite{Finelli:2000ya}.  In other words,
the power spectrum calculated at the end of multi-field inflation is
not necessarily the power spectrum observed in, say, the CMB data
because the post-inflationary dynamics (that is to say preheating
instabilities) can modify it.  In the same manner, it is tempting to
conjecture that the scalar bi-spectrum calculated at the end of
multi-field inflation will not necessarily be the same as the one
observed in the data~\cite{Enqvist:2004ey}. In particular, the
so-called consistency relations~\cite{Chen:2010xka}, which relate the
three-point and the four-point correlation functions (or,
equivalently, the corresponding dimensionless non-Gaussianity
parameters $\fnl$ and $\taunl$) might receive corrections in a
multi-field context due to metric preheating.  We are currently
investigating such issues.
 
\bibliography{biblio_preheat}
\end{document}